%% file: main.tex
\documentclass[sigconf,9pt]{acmart}
\copyrightyear{2026}
\acmYear{2026}
\setcopyright{cc}
\setcctype{by}
\acmConference[DAC '26]{63rd ACM/IEEE Design Automation Conference}{July 26--29, 2026}{Long Beach, CA, USA}
\acmBooktitle{63rd ACM/IEEE Design Automation Conference (DAC '26), July 26--29, 2026, Long Beach, CA, USA}
\acmDOI{10.1145/3770743.3804221}
\acmISBN{979-8-4007-2254-7/2026/07}
\usepackage{amsmath}
\usepackage{booktabs}
\usepackage{stfloats}
\usepackage{tikz}
\usetikzlibrary{decorations.pathreplacing}
\usepackage{arydshln}
\usepackage{enumitem}
\usepackage{multirow}
\usepackage{epsfig}
\usepackage{multicol,lipsum,xparse}
\usepackage{caption}
\usepackage{subcaption}
\usepackage{stmaryrd}
\usepackage{hyperref}
\usepackage{tikz}
\usepackage{algorithm}
\usepackage{algorithmicx}
\usepackage{algpseudocode}
\usepackage{pifont}
\usepackage{cleveref}

\usepackage{todonotes}
\usepackage{textcomp}

\usepackage{adjustbox}
\usepackage{xcolor}
\usepackage{booktabs, siunitx, pifont}
\newcommand{\cmark}{\ding{51}}
\copyrightyear{2026}
\acmYear{2026}
\setcopyright{cc}
\setcctype{by}
\acmConference[DAC '26]{63rd ACM/IEEE Design Automation Conference}{July 26--29, 2026}{Long Beach, CA, USA}
\acmBooktitle{63rd ACM/IEEE Design Automation Conference (DAC '26), July 26--29, 2026, Long Beach, CA, USA}
\acmDOI{10.1145/3770743.3804221}
\acmISBN{979-8-4007-2254-7/2026/07}

\begin{document}
\pagenumbering{roman}

\title{Autonomous Evolution of EDA Tools: \\ Multi-Agent Self-Evolved ABC}

\author{Cunxi Yu}

\affiliation{%
  \institution{NVIDIA Research}
  \institution{University of Maryland}
  \country{USA}
}


\author{Haoxing Ren}
\affiliation{%
  \institution{NVIDIA Research}
  \country{USA}
}


\newcommand{\fixme}[1]{\textcolor{red}{\small [~#1~]}}

\newcommand*\circled[1]{\raisebox{.4pt}
                    {\tikz[baseline=(char.base)]{
            \node[shape=circle,draw,inner sep=1pt, style={fill=white, text=black}, scale=0.75] (char) {\textbf{#1}};}}}

\begin{abstract}

This paper introduces the first \emph{self-evolving} logic synthesis framework,
which leverages Large Language Model (LLM) agents to autonomously improve
the source code of \textsc{ABC}, the widely adopted logic synthesis system. Our framework operates on the \emph{entire integrated ABC codebase},
and the output repository preserves its single-binary execution model and command interface. 
In the initial evolution cycle, we bootstrap the system using existing prior open-source synthesis components, covering
flow tuning, logic minimization, and technology mapping, but without manually injecting new heuristics. On top of this foundation, a team of LLM-based agents
iteratively rewrites and evolves specific sub-components of ABC following our ``programming guidance`` prompts under a unified correctness and QoR-driven evaluation loop.
Each evolution cycle proposes code modifications, compiles the integrated binary,
validates correctness, and evaluates quality-of-results (QoR) on \emph{multi-suite
benchmarks including ISCAS~85/89/99, VTR, EPFL, and IWLS~2005}. Through continuous feedback, the system discovers optimizations
beyond human-designed heuristics, effectively \emph{learning new
synthesis strategies} that enhance QoR. We detail the architecture
of this self-improving system, its integration with \textsc{ABC},
and results demonstrating that the framework can autonomously and
progressively improve EDA tool at full million-line
scale.

\end{abstract}
\maketitle

\input{01sec-intro.tex}

\input{02sec-background}

\input{04sec-approach}
\input{05sec-results}

\input{06-sec-conclusion}


\small
\bibliographystyle{ACM-Reference-Format}
\bibliography{ref.bib,cds.bib, ABCEVO}

\end{document}

%% file: 01sec-intro.tex
\section{Introduction}

Electronic Design Automation (EDA) tools form the computational backbone of modern hardware design. From logic synthesis and technology mapping to verification and physical design, these tools encapsulate decades of expert engineering, algorithmic innovation, and domain-specific heuristic design \cite{Stok1996BooleDozer,Mishchenko2010ABC}. Developing and maintaining such tools is extraordinarily challenging: the search spaces they operate on are combinatorial, algorithmic components interact in subtle ways, and new optimizations often require extensive re-engineering effort \cite{Sentovich1992SIS, Mishchenko2010ABC,murray2020vtr}. As a result, progress in EDA tool development is fundamentally constrained by the human-intensive, heuristic-driven nature of these systems.

Within this broader landscape, logic synthesis is a particularly representative domain. The \textit{ABC} framework has emerged as the de facto academic and industrial research platform for synthesis and verification \cite{Mishchenko2010ABC}. Over the past two decades, numerous research contributions, including cut-based mapping \cite{Mishchenko2007PriorityCuts, Fan2023AdaptiveMap}, Boolean rewriting \cite{Mishchenko2006Rewriting, Yu2017AIGrewrite}, flow-tuning algorithms \cite{Yu2018AutoFlow, Neto2023FlowTune}, etc., have been proposed as extensions to \textit{ABC}. However, most of these advances exist as external prototypes, loosely integrated scripts, or ad hoc binaries. Their capabilities often remain inaccessible within the monolithic \textit{ABC} codebase itself, due to the complexity of integrating new heuristics, the rigidity of existing subsystems, and the high development barrier of its low-level C implementation. Even within the built-in algorithms, many heuristics (e.g., cut selection, refactoring conditions, choice-node expansion, and cost estimation) are statically designed by experts and rarely revisited, despite the underlying NP-hard nature of their decisions \cite{Zhu2020RLlogic}. The result is a highly capable but difficult-to-evolve software system in which substantial QoR improvements may remain untapped.

Recent advances in code-generating AI, particularly Large Language Models (LLMs), provide a promising new perspective for automating EDA tool development \cite{ghose2026agentic, Yu2025SATLUTION,yao2026evoplace}. OpenAI’s Codex and DeepMind’s AlphaCode have demonstrated that LLMs can generate functioning code and even solve complex programming challenges at a competitive level \cite{Chen2021Codex, Li2022AlphaCode}. Google DeepMind's \textit{AlphaEvolve} demonstrated that an LLM-driven agent can iteratively refine algorithmic kernels and discover improvements beyond human-designed baselines \cite{Novikov2025AlphaEvolve}. Besides, NVIDIA's work \textit{SATLUTION} showed that LLM-based agents can evolve entire SAT solver repositories, producing solvers that surpassed human-engineered champions of the SAT Competition \cite{Yu2025SATLUTION}. These results suggest a provocative possibility: \textit{can EDA tools themselves with LLMs agents, with their complex algorithms and dense C/C++ implementations, become the target of autonomous code evolution?}

However, extending autonomous code evolution to a full EDA tool presents significant challenges. Existing efforts operate at limited scales: \textit{AlphaEvolve} focuses on isolated functions or small kernels (hundreds of lines of code) \cite{Novikov2025AlphaEvolve}, and \textit{SATLUTION} pushes repository-scale evolution to the order of tens of thousands of lines \cite{Yu2025SATLUTION}. In contrast, the \textit{ABC} logic synthesis and verification system contains more than \textbf{1.2 million lines of C code}\footnote{This line count was measured using the open-source \texttt{cloc} tool on our customized ABC repository. For comparison, the master branch of the original public ABC repository contains approximately 850K lines of code.}, spans over \textbf{4,000 source files}, and is organized across \textbf{four layers of hierarchical abstraction}. Its components are deeply interconnected, and even small changes create far-reaching effects throughout the tool. Techniques designed for modular or small-scale algorithm evolution do not directly apply: the sheer scale, cross-module dependencies, and multi-objective nature of synthesis (area, delay, QoR trade-offs) exceed the operational assumptions of prior LLM-based code-evolution frameworks.

This paper addresses these challenges by introducing a scalable,
multi-agent LLM-based framework for automatically evolving the
\textit{ABC} codebase itself. 
Specifically, we present a self-evolving logic synthesis framework that combines human domain expertise with agentic LLM-driven exploration. We begin with the baseline ABC tool augmented by several state-of-the-art extensions from recent research, thereby seeding the process with a strong initial design. On this foundation, our multi-agent system iteratively improves the tool, i.e., in each iteration, each agent proposes modifications to its respective module’s C code, the system verifies that the modified tool still produces correct logic transformations, and then evaluates the QoR of the resulting tool on a suite of benchmark circuits. Beneficial changes are retained and accumulated over many generations\footnote{We want to acknowledge that 
this work and the development of LLM-based agents would not be feasible
without the decades of foundational effort by the \textit{ABC} developers
and the broader logic synthesis community, whose open-source
implementations made this study possible.
}. Our contributions are summarized as follows:
\begin{itemize}[leftmargin=0pt]
\item We present the agentic LLM-based coding framework capable of
\textbf{autonomously evolving a multi-million-line EDA tool}, demonstrating
self-improving code generation at the full scale of the \textit{ABC}
logic synthesis system.
\item We introduce a \textbf{multi-agent evolution architecture} that
decomposes \textit{ABC}'s complex engineering structure into functional
subsystems, enabling specialized agents to collaboratively evolve shared
code under a unified QoR-driven objective.
\item We design a \textbf{formal correctness feedback} based on formal equivalence checking that
dramatically increases successful evolution steps while reducing LLM
token usage and computational cost.
\item We conduct a \textbf{comprehensive evaluation} across standard synthesis
benchmarks, showing that the evolved tool discovers improvements beyond
human-engineered baselines, demonstrating the feasibility of repository-scale
autonomous EDA tool development and its applicability to future EDA domains.
\end{itemize}

%% file: 02sec-background.tex
\section{Background}
\label{sec:background}

\subsection{Logic Synthesis and the \textsc{ABC} Framework}

Logic synthesis is a central component of the hardware design flow, responsible for transforming Boolean networks into technology-specific implementations for ASICs or FPGAs. It involves two tightly coupled phases: (1) \textit{logic optimization}, which restructures the network to reduce size or depth without altering correctness, and (2) \textit{technology mapping}, which binds the optimized logic to standard cells or LUTs. Both problems are NP-hard, and practical tools rely on multi-stage heuristics to balance quality-of-result (QoR) and runtime \cite{mishchenko2006rewrite, brayton2006resub}. \textsc{ABC} is the most widely used open-source platform for logic synthesis and verification research \cite{mishchenkoABC}, containing decades of algorithmic innovations across rewriting \cite{mishchenko2006rewrite}, resubstitution \cite{brayton2006resub}, refactoring \cite{mishchenko2011refactor}, and cut-based mapping \cite{Mishchenko2007PriorityCuts}. However, ABC’s effectiveness is inseparable from its extensive collection of hand-engineered heuristics: mapping cut filters, cost functions, refactoring thresholds, and flow scripts that were crafted through years of expert experimentation \cite{Yu2018AutoFlow, Neto2023FlowTune}.

Despite its long history, improving \textsc{ABC} has become increasingly difficult for several reasons. First, the codebase has grown to millions of lines across thousands of interdependent C files \cite{mishchenkoABC}. Small changes in one subsystem often propagate unpredictably to others due to shared data structures and tightly coupled invariants. As a result, even modifying a single heuristic typically requires deep technical knowledge of multiple modules and extensive manual testing, making development slow and error-prone. Second, QoR improvement itself has reached a saturation point under human-driven development. Modern flows contain many fixed decisions whose interactions are complex and circuit-dependent \cite{Yu2018AutoFlow}. Exhaustively searching this design space is infeasible: slight changes in pass ordering, local scoring functions, or mapping heuristics may yield significant QoR gains, yet these possibilities remain largely unexplored due to the prohibitive engineering effort required to evaluate them manually. Third, the pace of human engineering fundamentally limits progress. Creating, integrating, and validating heuristics requires days or weeks of expert effort for each iteration \cite{mishchenko2006rewrite}, whereas an automated system, once provided with correctness guards, can iterate continuously, generating and testing thousands of code-level modifications with no fatigue or attention limits. These challenges motivate the use of \textit{LLM-based coding agents} \cite{Chen2021Codex, Li2022AlphaCode} that operate directly on the \textsc{ABC} codebase. Such agents can rapidly explore the vast and largely untouched space of algorithmic variations, perform fine-grained code refinements, and autonomously search and implement new heuristics or optimization patterns that humans would be unlikely to design by hand.

\subsection{Self-Evolving Programming}

Recent advances in code-generating Large Language Models (LLMs) have enabled a new paradigm in which agents autonomously modify and improve software systems by iterating through cycles of code generation, execution, and feedback. Two prior efforts are most relevant to this direction. Google DeepMind’s \textit{AlphaEvolve} \cite{Novikov2025AlphaEvolve} demonstrated that LLM agents can refine and improve compact algorithmic kernels, typically only hundreds of lines of code, by repeatedly generating variants and evaluating their behavior. While groundbreaking, AlphaEvolve is limited to isolated, self-contained procedures with very few software dependencies or cross-module interactions. NVIDIA's recent work \textit{SATLUTION} \cite{Yu2025SATLUTION} extended the idea of autonomous code evolution to the repository scale in the domain of SAT solving. By coordinating planning and coding agents under strict correctness checks, including DRAT proof validation for UNSAT instances, SATLUTION was able to evolve full SAT solvers comprising tens of thousands of lines of C/C++ code. The resulting solvers surpassed human-engineered winners of the SAT Competition, showing that repository-level evolution is feasible when correctness can be rigorously enforced. 

However, applying these prior agentic frameworks directly to EDA tools is infeasible. The engineering and algorithmic complexity of logic synthesis systems such as \textsc{ABC} is far greater than that of modular SAT solvers or the small kernels used in AlphaEvolve. \textsc{ABC} contains more than {1.2 million lines} of C code distributed across {over 4000 source files} \cite{mishchenkoABC}, structured through multiple abstraction layers with substantial internal coupling. Further, unlike SAT solving \cite{Yu2025SATLUTION} or matrix multiplication \cite{smirnov2023mlalgorithms}, which is guided by a single scalar objective (runtime), logic synthesis optimization must jointly consider multiple metrics including area, delay, depth, node count, mapping cost, and correctness under technology constraints. These objectives are often in conflict, making reward design for autonomous evolution significantly more complex. Taken together, these factors reveal why existing LLM-based evolutionary frameworks cannot be naively applied to large-scale EDA tools. This motivates our framework, which introduces multi-agent coordination, extensive domain-guided structure, and a rigorous correctness and feedback loop to enable autonomous evolution at the full scale of the \textsc{ABC} logic synthesis system.

%% file: 04sec-approach.tex
\section{Approach}
\label{sec:approach}

\begin{figure*}[t]
\centering
\includegraphics[width=0.9\linewidth]{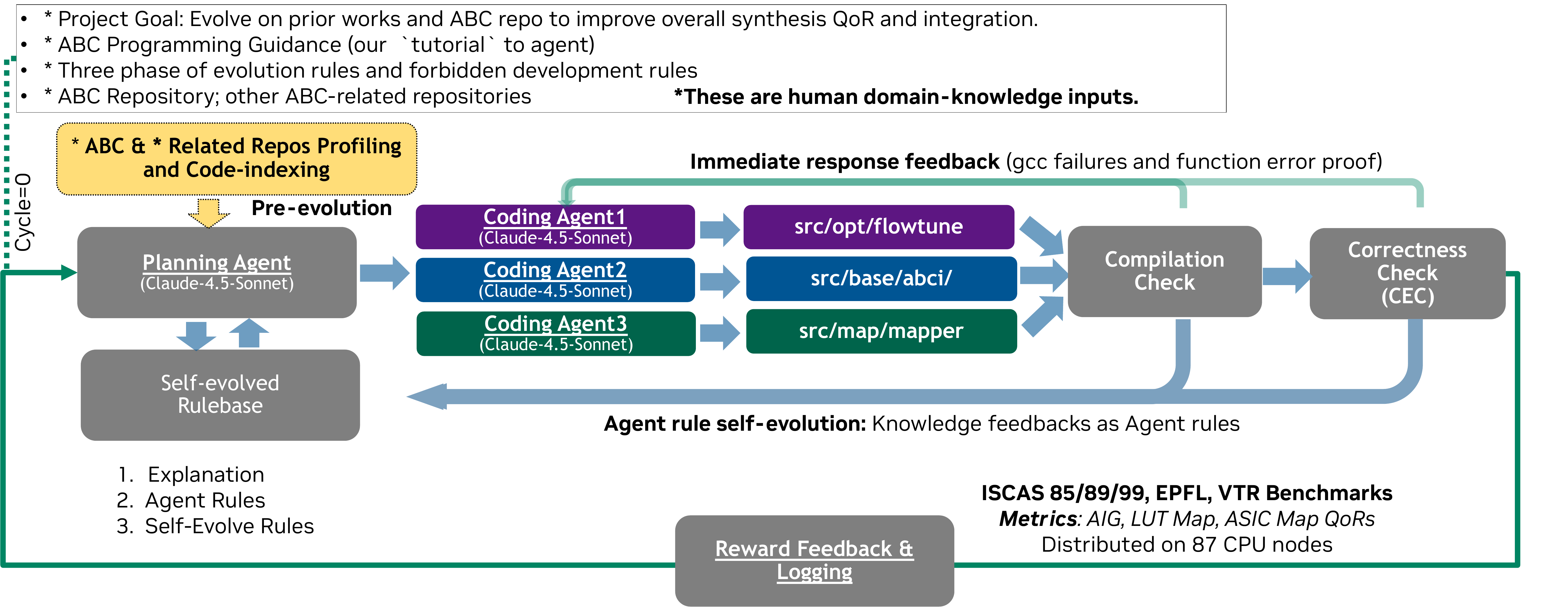}
\caption{Overview of the multi-agent self-evolving framework for \textsc{ABC}. 
Specialized LLM agents evolve distinct subsystems (flow optimization, core algorithms, and mapping), 
with each iteration undergoing compilation, formal CEC verification, and full QoR evaluation. 
A planning agent coordinates global decisions, a coding agent implements edits, 
and all agents follow a shared rulebase and unified evaluation pipeline to enable coordinated, correctness-preserving improvements.}

\label{fig:overview}
\end{figure*}


\subsection{ Knowledge Bootstrapping}

To bootstrap our framework, we performed a comprehensive prior knowledge and codebase exploration. In this \textbf{pre-evolution stage} (Figure \ref{fig:overview}), an autonomous agent surveyed numerous open-source projects and research prototypes in three key categories, i.e., \emph{flow tuning} \cite{yu2020flowtune,Neto2023FlowTune,Zhu2020RLlogic,zhu2020rlsynthesis}, \emph{technology mapping algorithms} \cite{Mishchenko2007PriorityCuts}, and \emph{technology-independent optimization algorithms \cite{mishchenko2006rewrite, brayton2006resub, mishchenkoABC, mishchenko2011refactor, Mishchenko2007PriorityCuts}}. The agent evaluated each repository for quality, extensibility, and potential synergy with an AI-driven self-improvement loop (i.e., self-evolution), with particular focuses on full ABC integration. The agent ranked the candidates and selected the most promising codebases in each category as starting points for our approach. Below we summarize the outcomes of this preparation phase, highlighting the chosen frameworks and their advantages.

\textbf{Flow Tuning:} The agent began by pre-profiling several open-source flow-tuning frameworks and analyzing their engineering structures, with particular attention to how tightly each integrates with the \textsc{ABC} codebase. Although many recent learning-based flow optimizers treat \textsc{ABC} as a black-box binary and rely on external ML frameworks (e.g., PyTorch) to drive pass selection, these approaches provide limited opportunity for direct code evolution inside \textsc{ABC}. In contrast, \textit{FlowTune} is implemented as a fully integrated synthesis command within \textsc{ABC}, with a clean internal interface and a modular directory structure. Recognizing this advantage for repository-scale evolution, the agent selected FlowTune as the backbone for flow-level exploration. Furthermore, the agent also recommended flow tuning coding interface and structure in ABC at \texttt{src/opt/flowtune}, describing how new modules should be added to the build system and automatically created appropriate updates to \texttt{module.make}, following FlowTune's original implementation. While FlowTune provided the practical integration substrate, the agent also absorbed conceptual insights from external work, such as GNN-based flow selection  \cite{zhu2020rlsynthesis}, to inform possible heuristic modifications during later evolution cycles.

\textbf{Technology Mapping:} For technology mapping, the agent systematically surveyed existing methods, including ABC's built-in mappers, ML-augmented cut-ranking methods \cite{liu2023aimap, neto2021slap}, and mapping tuning techniques \cite{liu2024maptune,liu2026maptune}. A detailed analysis by the agent of external approaches revealed significant integration challenges, primarily due to their reliance on external learning frameworks incompatible with the in-repo code structure. Consequently, the agent determined that direct reuse of these external implementations was not recommended. Instead, the agent leveraged the architectural knowledge gleaned from examining the codebase. This examination effectively highlighted the core regions within $\textsc{ABC}$'s mapper—specifically where cut enumeration, pruning, and cost scoring occur. By identifying which internal $\textsc{ABC}$ functions implement cut-filtering logic and priority heuristics, the agent accurately localized the precise injection points for new heuristics, pruning criteria, or cost-model refinements. Thus, the external implementations influenced the mapping agent not through direct code integration, but by providing the structural guidance necessary to bootstrap the evolution of the existing $\textsc{ABC}$ mapping routines.

\textbf{Technology-Independent Optimization:} The largest and most fertile region for self-evolution comes from technology-independent optimization, where \textsc{ABC} contains extensive AIG rewriting, refactoring, and resubstitution code. Here the agent studied both internal implementations and external research prototypes such as \textit{Orchestration} \cite{li2024dag}. By cross-referencing these implementations with \textsc{ABC}’s own optimization engine, the agent gained a rich understanding of the various heuristics, thresholds, and transformation operators available for modification. At the end of profiling, the agent prompts to work on orchestration in combination with flow tuning to add additional synthesis command interface to improve overall flow.



\subsection{Agent Setups and Agent Roles}
Rather than relying on a single monolithic agent attempting to modify the entire \textsc{ABC} codebase, our framework employs multiple \emph{specialized} agents, each responsible for a well-scoped subsystem. 
We instantiate three domain-aligned coding agents. The \textbf{Flow Agent} operates on the flow scheduling and pass-orchestration layer of \textsc{ABC}, initially seeded with the FlowTune C-based implementation. 
The \textbf{Mapper Agent} targets the technology-mapping subsystem; its initial working directory is cloned directly from the vanilla \textsc{ABC} mapper (\texttt{src/map/mapper}), while its initial planning was informed by SLAP’s cut-selection formulation, enabling the agent to localize and modify cut pruning, enumeration, and cost-scoring heuristics. Finally, the \textbf{Logic Minimization Agent} focuses on technology-independent optimization at \texttt{src/base/abci/}. 

All agents share a unified objective: improve the tool’s overall QoR without altering functional semantics. Each agent is realized through Claude Sonnet~4.5 queries: a \emph{Planning} that formulates subsystem-specific modification strategies, and a \emph{Coding Agent} that implements those strategies by generating diffs and editing the repository. 
Importantly, \emph{only cycle~0 is guided by human-provided context} as shown in Figure \ref{fig:overview}, including repository profiling, structured Markdown tutorials, and subsystem initialization. After initialization, all subsequent evolution is autonomously planned and executed by the agents. Human intervention is limited strictly to safety triggers, only occurring after more than ten consecutive attempts fail due to compilation errors or equivalence-checking violations. 

\subsection{Evolutionary Iteration Workflow}
Each evolutionary iteration follows a structured sequence:

\begin{enumerate}[leftmargin=0pt]

\item \textbf{Planning}:  
At the center of the system is a single \textbf{Planning Agent}, instantiated using Claude Sonnet~4.5, which oversees all stages of the self-evolution process. The planner is responsible for global coordination: it interprets the QoR feedback from previous cycles, aggregates hypotheses and change proposals from each coding agent, and determines which subsystem should be evolved next. In cycle~0, the planning agent is provided with a structured and comprehensive introduction to the \textsc{ABC} framework. Specifically, we supply: (i) a repository-wide profiling generated by the agent itself in pre-evolutio stage, summarizing the directory hierarchy, build system, and module interactions; (ii) a highly structured Markdown tutorial, authored by us, detailing how to add new functions, register new commands, modify existing data structures, and interact with the \textsc{ABC} synthesis and mapping APIs; (iii) distilled explanations of common synthesis flows derived from \texttt{abc -h}, online documentation, and our own understanding of \textsc{ABC}’s internal architecture. We found that providing these materials in Markdown, rather than natural language, significantly improved agent reliability, as the structured format encourages procedural reasoning and reduces hallucination. For cycle $\geq1$, the planner agent takes the feedback metrics together with the multi-agent hypotheses and code changes as input, and autonomously plans the next evolution step.

\item \textbf{Coding Phase}:  
The three coding agents operate on distinct and non-overlapping subsystems of the \textsc{ABC} codebase, as shown in Figure~\ref{fig:overview}. The first agent, responsible for \emph{optimization flow evolution}, works within the flow-scheduling and pass-orchestration layer, primarily interacting with the FlowTune-integrated module located under \texttt{src/opt/flowtune/}. Its role is to evolve pass-selection heuristics, stopping criteria, and conditional flow steps, ensuring that any modifications are local to the FlowTune module to avoid interference with core \textsc{ABC} commands. The second agent, the \emph{mapping agent}, operates in the technology mapper subsystem (\texttt{src/map/mapper/}), where it refines heuristic parameters for cut enumeration, pruning, and cost scoring. Although SLAP itself is not directly integrated, its structure guided the agent in identifying which mapping routines were safe to modify and which must remain untouched for compatibility with \textsc{ABC}'s existing mapping engine. The third agent focuses on \emph{technology-independent logic optimization}, working within existing module \texttt{src/base/abci/}, and the orchestration extensions. All three agents apply changes only within their designated directories, preventing conflicts with other subsystems and preserving \textsc{ABC}’s architectural invariants. To ensure full compatibility with the original \textsc{ABC} build system, each agent-generated module adheres to the repository’s \texttt{module.make} structure and Makefile conventions; the planning agent explicitly learned these conventions during the cycle~0 repository-profiling phase. As a result, newly introduced source files or modified modules integrate seamlessly into the existing Makefile hierarchy without overwriting or conflicting with core \textsc{ABC} functionality.

\item \textbf{Compilation and Correctness Pre-Checks}:  
The modified repository is compiled. If compilation errors occur, the failure logs are returned to the coding agent, which enters a rapid “self-debugging” loop until the build succeeds. Once compilation passes, basic sanity checks are run. Crucially, we perform \emph{formal equivalence checking (CEC)} using \textsc{ABC}’s combinational equivalence engine. Because our system does not apply retiming or sequential logic changes, sequential benchmarks can also be validated through CEC via standard unrolling or internal correspondence. Any mismatch, counterexample, or unintended logic change causes immediate rejection of the iteration and produces corrective feedback for the next cycle.

\item \textbf{Benchmark Evaluation and Feedback Integration}:  
Once correctness is verified, the evolved \textsc{ABC} version is evaluated using a large-scale distributed workflow designed to maximize feedback density per iteration. We deploy the evaluation across a cluster of 87 CPU nodes, enabling all benchmarks and all synthesis flows to be executed simultaneously. For each evolution cycle, every design is synthesized under eight distinct flows (full ablation suite) to expose how the modified algorithms behave across diverse optimization regimes. These flows include variations of optimization, mapping, buffering, and gate-sizing strategies, all implemented using the ASAP7 7\,nm PDK. This exhaustive evaluation produces a high-resolution characterization of each iteration's behavior.

For each benchmark and flow, we collect both final QoR metrics, post-mapping area, STA timing analysis, and critical path delay, and a rich set of intermediate structural signals extracted directly from \textsc{ABC} (Table \ref{tab:qor-metrics}). These include AIG node count, depth, edge count, mapper-level area and delay estimates before buffering and sizing, internal traversal statistics, and per-pass structural deltas. The combined feedback from these metrics forms a multi-dimensional performance profile that captures both early-stage structural improvements during optimization and end-of-flow QoR outcomes.

From this data, the system computes a scalar reward and a detailed QoR delta vector. Improvements are incorporated into the global “champion” version of the tool; regressions trigger an immediate rollback of the affected subsystem. Partial improvements, for example, a mapping enhancement that reduces depth but slightly increases pre-buffer area, can be conditionally preserved when they exhibit synergy potential with evolving components in other subsystems, as determined by the planner.

\begin{table}[t]
\scriptsize
\centering
\caption{QoR metrics collected during each evaluation cycle. The top group lists primary optimization targets, while the lower group lists auxiliary metrics that guide evolution.}
\vspace{-4mm}
\label{tab:qor-metrics}
\begin{tabular}{l l}
\toprule
\textbf{Metric Category} & \textbf{Description} \\
\midrule
\multicolumn{2}{l}{\textit{Primary (Targeted) Metrics}} \\
\midrule
STA Timing (Worst Slack)  
    & STA Timing Results (ASAP7 7\,nm) \\
Post Buffer/Sizing Area  
    & Area after buffering and gate sizing \\
\midrule
\multicolumn{2}{l}{\textit{Intermediate Structural and QoR Metrics, and non-direct metrics in Feedback}} \\
\midrule
AIG Metrics  
    & \#nodes, \#edges, \#depth \\
Mapper Area  
    & Pre-buffer/sizing area \\
Mapper Delay Estimate  
    & Logic-level delay  \\
Cut Enumeration Statistics  
    & Number of cuts, pruned cuts, cut sizes \\
Structural Deltas per Pass  
    & Change in size/depth per optimization step \\
LUT Map Results (additional flow feedback)  
    & \#LUTs, depth \\
\bottomrule
\vspace{-5mm}

\end{tabular}
\end{table}

\item \textbf{Self-Evolving Rulebase}:  
A central contribution of our framework is the \emph{self-evolving rulebase} that governs multi-agent coordination and constrains code modification. Beyond correctness safeguards, the rulebase provides subsystem-specific policies that guide how each agent may alter heuristics, thresholds, cost models, and traversal strategies. During evolution, the planner continuously evaluates whether these rules are overly restrictive or misaligned with emergent patterns in QoR feedback. When a rule consistently blocks beneficial edits, the planner may propose a controlled relaxation or refinement of that rule, subject to global safety constraints. This dynamic rule evolution enables the system to shift naturally from conservative behavior in early cycles, where the primary objective is stability, to progressively more exploratory and structural modifications in later cycles. 

\end{enumerate}

\subsection{Correctness guided by Equivalence Checker}

 A single incorrect rewrite rule, mapping-cost update, or AIG manipulation can silently corrupt the synthesized netlist and mislead the evaluator with false QoR improvements. As observed in SATLUTION \cite{Yu2025SATLUTION}, such failures can cause large-scale wasted LLM cycles, repeated rollback loops, and degraded convergence. To prevent this, we leverage SATLUTION formal feedback concept to employ \emph{formal combinational equivalence checking (CEC)} after every code modification and before any QoR evaluation occurs. Because the evolved algorithms do not alter retiming or sequential behavior, CEC is sound for all benchmarks: combinational circuits are checked directly, and sequential circuits are validated via a single-frame bounded-model formulation (BMC depth is 1) that proves equivalence of the combinational logic feeding each register. We use the ``\texttt{cec}`` and ``\texttt{dsat}`` commands in the ABC framework to perform CEC on all designs before and after optimization with the evolved code at each iteration. Any mismatch or SAT-derived counterexample immediately terminates the iteration and provides corrective feedback to the next planning cycle. 

%% file: 05sec-results.tex
\section{Results and Discussion}
\label{sec:results}

\textbf{Expeimental Setups:} All experiments are executed on a distributed cluster of 87 AMD EPYC CPU nodes, allowing every benchmark and every synthesis flow (including CEC) to be evaluated in parallel at each evolution iteration. The multi-agent system is implemented using the enterprise-version \textbf{Cursor} environment with all agents instantiated as Claude 4.5 Sonnet models. For each cycle, the evolved \textsc{ABC} version is evaluated on eight synthesis flows using the ASAP7 7\,nm library, covering the full logic optimization, mapping, buffering, gate sizing, and static timing analysis stages. Benchmarks include ISCAS’85 and ISCAS’89, ITC’99 \cite{davidson1999itc}, EPFL \cite{amaru2015epfl}, VTR DSP \cite{murray2020vtr}, and a collection of arithmetic blocks. 

\begin{table}[t]
\footnotesize
\centering
\caption{Ablation on evolution across FlowTune (FT), AIG synthesis (Orch), and mapping (Map). Avg QoR is normalized to the baseline configuration (Vanilla FT + Vanilla Orch + Vanilla Map = 1.00). Lower is better.}
\label{tab:ablation-evo}
\setlength{\tabcolsep}{3.5pt}
\begin{tabular}{lccccccS[table-format=1.3]}
\toprule
\multirow{2}{*}{Config.} &
\multicolumn{2}{c}{FT} &
\multicolumn{2}{c}{Orch} &
\multicolumn{2}{c}{Map} &
{QoR} \\
\cmidrule(lr){2-3} \cmidrule(lr){4-5} \cmidrule(lr){6-7}
& Vanilla & Evo & Vanilla & Evo & Vanilla & Evo & {} \\
\midrule
Vanilla ABC        & - & - & - & - & - & - & 1.213 \\
Vanilla all        & \cmark &  & \cmark &  & \cmark &  & 1.000 \\
+ Evo FT           &   & \cmark & \cmark &  & \cmark &  & 0.962 \\
+ Evo Orch         & \cmark &  &   & \cmark & \cmark &  & 0.957 \\
+ Evo Map          & \cmark &  & \cmark &  &   & \cmark & 0.988 \\
+ Evo FT + Orch    &   & \cmark &   & \cmark & \cmark &  & 0.924 \\
+ Evo FT + Map     &   & \cmark & \cmark &  &   & \cmark & 0.939 \\
+ Evo Orch + Map   & \cmark &  &   & \cmark &   & \cmark & 0.942 \\
\textbf{All Evo}   &   & \cmark &   & \cmark &   & \cmark & \textbf{0.917} \\
\bottomrule
\end{tabular}
\end{table}

\subsection{Self-Evolved ABC Evaluation}
Table~\ref{tab:ablation-evo} summarizes the averaged quality of results for all combinations of evolved and vanilla subsystems, normalized to the baseline configuration that uses vanilla FlowTune \cite{yu2020flowtune,Neto2023FlowTune}, vanilla AIG Syn, and vanilla Map. All evolved code are direct ABC commands along with all the other ABC vanilla commands. AIG Syn refers to evolved versions of \texttt{(rewrite, resub, refactor, orchestrate)}$_x$, where the flow sequence will be optimized by \texttt{flowtune$_x$} command. Note that $x$ is the iteration cycle count. Lower values indicate improvement. Evolving a single subsystem yields modest benefits, for example Evo FlowTune alone reduces the normalized QoR from 1.000 to 0.962 and Evo AIG Syn alone to 0.957, while Evo Map alone provides a smaller gain at 0.988. More substantial improvements emerge when two subsystems co-evolve. Evo FlowTune plus Evo AIG Syn achieves a QoR of 0.924, and Evo FlowTune plus Evo Map reduces it to 0.939. Evo AIG Syn plus Evo Map performs similarly, reaching 0.942. The strongest gains arise when all three subsystems evolve jointly, achieving the best result of 0.917, corresponding to an overall improvement of approximately 8.3 \% relative to the vanilla integrated baseline. These differences reflect the complementary roles of the subsystems: the FlowTune agent enhances upstream structural simplification, the Orchestrate agent reduces mid-level logic redundancy and depth, and the Mapping agent introduces depth-sensitive tie-breaking rules that mitigate mapped delay. Their combined evolution produces a consistently superior end-to-end flow.

Across all benchmark suites the fully evolved system improves both static timing and area–delay product. Worst negative slack improves by approximately 8 to 9 \% on average relative to the vanilla integrated baseline, with several EPFL arithmetic circuits exhibiting improvements between 12 and 15 \%. The area–delay product reduction of roughly 8.3 \% aligns with the 0.917 normalized value in Table~\ref{tab:ablation-evo}. Intermediate structural metrics further corroborate these trends. AIG node counts decrease by 3 to 8 \% on arithmetic-heavy designs, and post-mapping depth decreases by 4 to 6 \% due to depth-aware heuristics introduced autonomously by the mapping agent. These patterns indicate that the multi-agent framework does not merely tune constants but instead learns to restructure the synthesis process at several algorithmic layers, producing meaningful and consistent QoR improvements across diverse circuit families.

 
\vspace{-5mm}
\subsection{Computational Cost and Token Usage}
The majority of tokens in our framework are spent during the initial profiling of the \textsc{ABC} codebase, which accounts for 68\% of total LLM usage, while an additional 11\% is devoted to profiling external codebases during initialization. Only the remaining 21\% of tokens correspond to the evolution cycles themselves, demonstrating that once the system is fully primed with repository knowledge, subsequent iterations incur relatively low marginal token cost. With 87 CPU nodes running in parallel, each full iteration, including all benchmarks across all eight flows and full CEC, completes in approximately 2-3 hours, enabling rapid turnaround and allowing the planner to incorporate high-quality feedback at scale. Token usage per evolution cycle corresponds to an estimated cost of \$60--\$80 depending on the number of repairs required by the coding agents. Because full-design equivalence checking prevents faulty versions from progressing to evaluation, more than 90\% of tokens are spent on semantically valid code states, substantially reducing waste and stabilizing the learning dynamics. Moreover, due to the scale of ABC, we notice that the majority of the token spent is at pre-evolution stage, i.e., profiling ABC vanilla repository and other codebase. For the composition of generated content, where 45\% of all produced artifacts correspond to C source code (87{,}749 lines in total), and the remainder consists of Markdown, bash, Python, and log. The initial code-profiling stage represents another significant portion, with approximately \$1{,}400 spent at initialization. The total token cost of the full system is approximately \$2{,}400.

\subsection{Self-Evolved Code Quality and Discussion}

\begin{figure}
    \centering
    \includegraphics[width=1\linewidth]{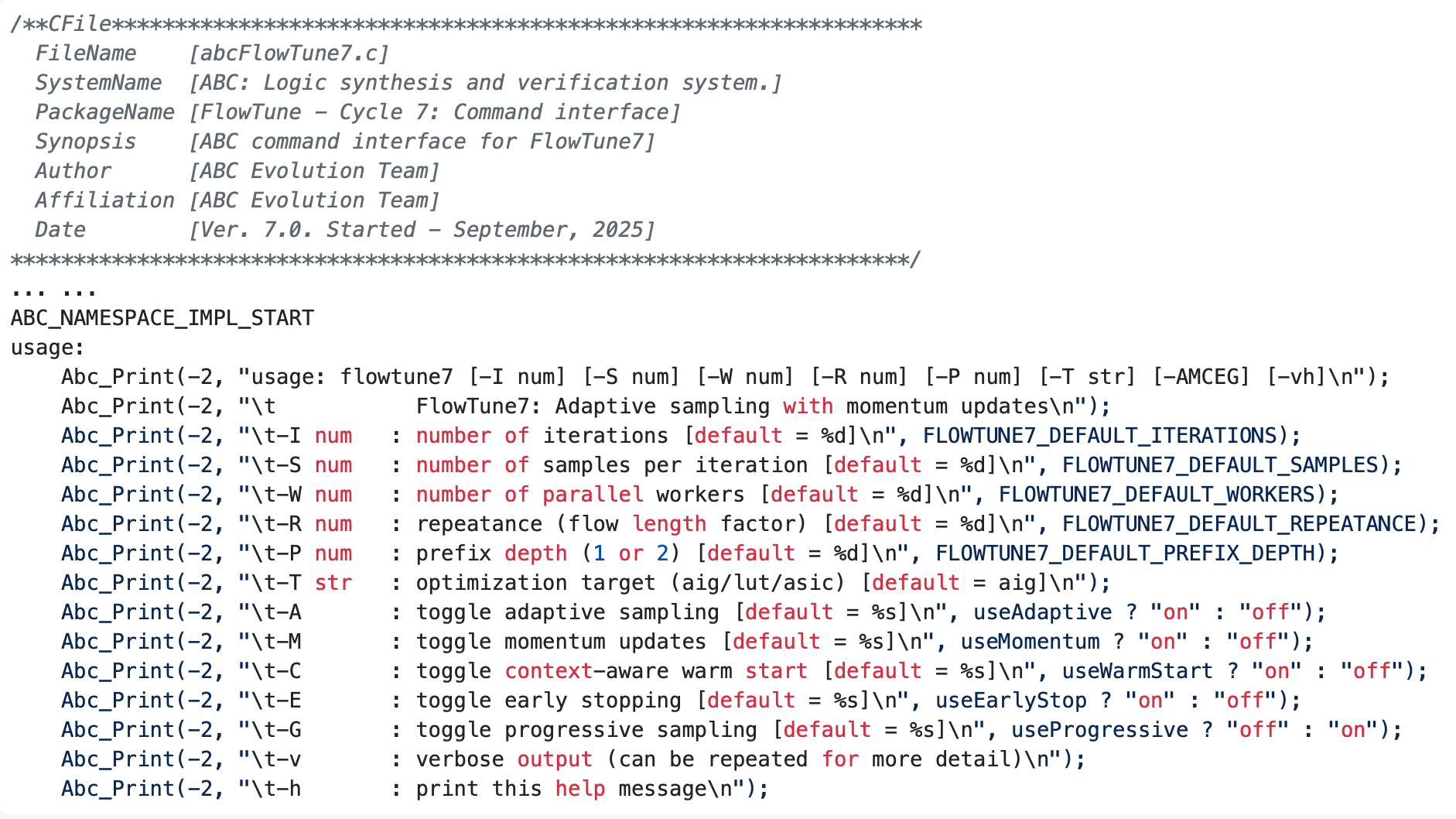}
    \caption{Automatically generated \texttt{abcFlowTune7.c} module partial code produced during evolution (cycle 7).}
    \vspace{-5mm}
    \label{fig:code7}
\end{figure}

An additional and noteworthy outcome of this study is the consistently high code quality produced by the evolved agents. Despite exposure to heterogeneous external research repositories during initialization, the agents overwhelmingly converge to the native \textsc{ABC} coding style when generating new C code. The patches they produce closely match \textsc{ABC}’s formatting, naming conventions, commenting structure, and macro organization with striking fidelity. For example, the automatically generated \texttt{abcFlowTune7.c} module shown in Figure~\ref{fig:code7} adopts the exact header layout, namespace structure, and \texttt{Abc\_Print} usage patterns of hand-written \textsc{ABC} components, including correctly reproduced parameter help strings, indentation, and macro-driven defaults. 

Beyond formatting, the agents excel at refining algorithmic directions that already have structural precedent within \textsc{ABC}. They reliably adjust overlooked thresholds, enhance depth-aware scoring, and introduce conditional heuristics that improve QoR. In contrast, attempts to introduce entirely novel algorithmic constructs, without anchors in existing invariants, tend to fail more often, typically due to compilation errors, segmentation faults, or subtle correctness violations that cannot be self-debugged. This pattern highlights both the power and current limitations of LLM-driven evolution: the framework is highly effective at amplifying and recombining human-curated priors into stronger heuristics but is less reliable at proposing fully new paradigms without scaffolding.

%% file: 06-sec-conclusion.tex
\vspace{-2mm}
\section{Conclusion}
\vspace{-1mm}
 This work demonstrates that \emph{self-evolving} agentic framework is capable of autonomously improving EDA tool. By coupling LLM-based multi-agent programming with correctness-preserving compilation and QoR-driven evaluation, the system effectively learns new algorithmic strategies directly over the entire \textsc{ABC} codebase. 
 These results highlight a new paradigm for EDA: repository-scale, agentic, self-improving tool development, possibly opening a path toward future EDA tool development. \textit{However, we would also like to emphasize that, at the current stage, agents perform well only when they are guided by sufficient domain knowledge. More importantly, the success of agents on the EDA problems studied in this work is not created in isolation. We believe it is enabled by decades of foundational contributions from the EDA community, whose pioneering efforts established the algorithms, tools, and knowledge base on which these agents now build.}

\textbf{Acknowledgment.} The authors would like to thank Prof.~Zhiru Zhang and his students for their valuable feedback and insightful discussions.

\newpage